\documentclass[12pt]{article}

\usepackage[margin=1in]{geometry}

\usepackage{natbib}
\usepackage[colorlinks=true, allcolors=blue]{hyperref}
\usepackage{amsmath}
\usepackage{bbm}
\usepackage[title]{appendix}
\usepackage{xcolor}
\usepackage{rotating} 
\usepackage{multirow}
\usepackage{threeparttable}
\usepackage{subcaption}
\usepackage{xr-hyper} 
\newcommand{\bx}{\boldsymbol{x}}
\newcommand{\bX}{\boldsymbol{x}}
\newcommand{\by}{\boldsymbol{y}}
\newcommand{\bt}{\boldsymbol{t}}
\newcommand{\bxi}{\boldsymbol{\xi}}
\newcommand{\bth}{\boldsymbol{\theta}}

\newcommand{\ths}{\theta^\star}
\newcommand{\hth}{\widehat{\theta}}

\newcommand{\tg}{\widetilde{g}}
\newcommand{\ts}{\widetilde{s}}
\newcommand{\tXi}{\widetilde{\Xi}}
\newcommand{\tP}{\text{Poisson}}

\newcommand{\mU}{\mathcal{U}}
\newcommand{\wmU}{\widetilde{\mathcal{U}}}
\newcommand{\wL}{\widetilde{L}}
\newcommand{\E}{\mathbbm{E}}
\newcommand{\bone}{\mathbbm{1}}
\newcommand{\mT}{\mathcal{T}}
\newcommand{\mD}{\mathcal{D}}

{
	\title{\bf Monitoring Adverse Events Through Bayesian Nonparametric Clustering Across Studies
 	}
    \author{
Shijie Yuan\thanks{Department of Statistics and Data Sciences, The University of Texas at Austin, Austin, TX 78712, USA; Corresponding email: sj.yuan@utexas.edu}\ ,
Kevin Roberts\thanks{Pfizer Inc.; Email: kevin.roberts@pfizer.com}\ ,
Noirrit Kiran Chandra\thanks{Department of Mathematical Sciences, The University of Texas at Dallas, Richardson, TX 75080, USA; Email: noirritchandra@gmail.com}\ ,
Yuan Ji\thanks{Department of Public Health Sciences, The University of Chicago, Chicago, IL 60637, USA; Email: koaeraser@gmail.com}\ ,
and \\
Peter M\"uller\thanks{Department of Mathematics, The University of Texas at Austin, Austin, TX 78712, USA; Email: pmueller@math.utexas.edu}\ 
 	}
	\date{\today}
}

\begin{document}
\maketitle
\thispagestyle{empty}

\newpage
\thispagestyle{empty}
\begin{abstract}
We introduce a Bayesian nonparametric inference approach for aggregate adverse event (AE) monitoring across studies.
The proposed model  seamlessly integrates  external data from historical trials to define a relevant background rate  and accommodates varying levels of covariate granularity (ranging from patient-level details to study-level aggregated summary data). Inference is based on a covariate-dependent product partition model (PPMx). A central element of the model is the ability to group experimental units with similar profiles. 
We introduce a pairwise similarity measure,  with which  we   set up a random partition of  experimental units with comparable covariate profiles, thereby improving the precision of AE rate estimation.
Importantly,  the  proposed  framework supports real-time safety monitoring under blinding with a seamless transition to unblinded analyses when indicated. Using one case study and simulation studies, we demonstrate the model’s  ability to detect safety signals and assess risk under diverse trial scenarios. 

\end{abstract}

\clearpage\newpage
\setcounter{page}{1}
\section{Introduction}

We introduce an inference model with multi-resolution structure and incorporating external data sources to monitor aggregate adverse event (AE) rates  in clinical trials.
Building on this model we propose inference to compare AE incidence rates in  an ongoing  
clinical trial to  a background rate that is constructed using external data, such as  historical trials. The model adjusts for the heterogeneity of different studies and cohorts by accounting for diverse covariates, such as treatment, dose, age, indications, and other relevant characteristics. To integrate data with varying levels of granularity  (i.e., resolutions),   ranging from 
 patient-level information to study-level aggregated summaries,
the proposed model defines experimental units as the fundamental elements for analysis by study, treatment arms or other cohorts. 
 These units may correspond to distinct combinations of covariates in patient-level data,  cohorts  defined as   subsets of patients  based on their  
 (marginal) covariate levels, or an entire study, depending on the levels of aggregation in the available AE rates for each study.
We implement borrowing of strength across different data sources by  grouping units with similar covariate profiles.  The cluster arrangement is treated as a random element, and all inference is marginalized w.r.t. the uncertainty in clustering. 
This allow us  to enhance AE monitoring through the integration of external data and robust clustering. 
Methodologically, the proposed model implements regression (of AE rates) with varying levels of covariate summaries, including variable dimensional covariate vectors.
The proposed approach formalizes the decision process outlined in  US  FDA guidance \citep{us2021sponsor} as based on decision boundaries in a coherent underlying posterior  probability model on  all unknown quantities, including, in particular, the relevant background rate.

Randomized controlled trials (RCTs) are widely regarded as the gold standard for evaluating the efficacy and safety of treatments by randomly assigning participants to treatment or control arms,  allowing investigators to attribute any difference in outcomes to only treatment assignments. 
Safety monitoring and reporting are integral components of RCTs, ensuring the protection of participants and  assuring meaningful inference on efficacy outcomes (lest an improvement in efficacy could be clinically meaningless due to unacceptable AE rates). 
The U.S. FDA underscores the importance of robust safety assessment mechanisms and encourages sponsors to conduct aggregate safety analyses, particularly for anticipated  AEs  
and increased rates of serious suspected adverse reactions \citep{us2021sponsor}. 
Anticipated  AEs are expected  
to occur in the study population due to the underlying disease or background factors, but not necessarily related to the investigational drug.
Serious suspected adverse reactions are serious adverse events for which there is a reasonable possibility that the investigational drug caused the event.
Figure~\ref{fig:flowchart} presents a redrawn version of the FDA’s aggregate analysis decision framework \citep{us2021sponsor}, included here for clarity and consistency with the regulatory reporting context. 
To meet these regulatory expectations and address the challenges posed by increasingly complex RCT designs,  approaches to  advanced safety monitoring are essential.  Ideally, safety monitoring should   integrate real-time data with historical trials, enable enhanced signal detection and improved safety monitoring,  and overall foster  a more comprehensive understanding of treatment safety \citep{barnes2021risk}.

There is a fast growing literature on the  topic  of incorporating historical trials in safety monitoring.
\citet{cai2010meta} discuss the use of meta-analysis methods for rare adverse events,  introducing  Poisson random effects to evaluate treatment-related risks. 
\citet{french2012using}  propose  a Bayesian methodology for early monitoring in clinical trials by incorporating historical control data, providing calibrated decision-making criteria to balance early termination risks and safety.
\citet{mukhopadhyay2018bayesian}  introduce  a two-step Bayesian method for blinded safety data monitoring. Their approach involves screening for safety signals using posterior probabilities based on a Poisson model for AE counts  and conducting sensitivity analyses based on historical data. 
\citet{brock2023bayesian} present a Bayesian approach to safety signal detection in blinded randomized controlled trials. This method integrates meta-analytic predictive priors derived from historical data with a Bayesian model to monitor adverse event probabilities without unblinding trials. 
See \citet{phillips2020statistical} for a review of statistical methods for analyzing AE data in practical RCTs. Notably, ten methods in this review directly consider historical or external data by leveraging Bayesian principles,  highlighting  the benefit of incorporating prior knowledge to improve decision-making, especially in cases of rare or emerging AEs.

Although most of these papers account for heterogeneity across historical datasets, they generally do not incorporate  specific, 
 features  of the historical trials,  but instead  impose a  uniform strength of information borrowing across trials. 
Building on a covariate-dependent product partition model (PPMx) introduced by \citet{muller2011product} and \citet{page2022clustering}, we  set up covariate-dependent random partitions (of experimental units), which allow us to increase 
the probability that experimental units with similar covariates are grouped together for increased levels of borrowing strength. 
In particular, we build on a pairwise similarity coefficient \citep{gower1971general}, proposed in \citet{dahl2017random}, to quantify the closeness of individual covariate vectors. 
In the context of comparing clinical studies or patient subpopulations, noting that pairwise similarity of experimental units provides a more interpretable framework. In this manuscript we introduce a PPMx based on cluster-level similarity functions based on average pairwise similarity of units. Furthermore, our proposed framework is designed to seamlessly transition between blinding and unblinding during safety monitoring, ensuring both the preservation of trial integrity and the timely identification of critical safety concerns. 

In addition to leveraging historical trials and covariate-dependent clustering for improved safety monitoring, our proposed framework also highlights the construction and application of background rates—defined as the expected incidence of specific adverse events in the study population independent of drug exposure. These rates provide the critical context to determine whether observed AEs are drug-related or naturally occurring \citep{us2021sponsor}.
Generally, background rates are derived by incorporating external data—for example, historical trials, epidemiological studies, or electronic health records \citep{hennessy2011pharmacoepidemiology}.
Within our proposed framework, the background rates  are systematically estimated in model-based inference  accounting for the similarity between units in the current study and those from historical datasets.

The remainder of this paper is organized as follows. In Section \ref{sec:motivating_app}, we present the motivating application, focusing on the context and challenges that drive the development of our framework. Section \ref{sec:inference_model}  describes   the proposed inference model, including its formulation and the integration of covariates for clustering. Section \ref{sec:decision_problem} addresses the decision problem, outlining the criteria and methods for safety monitoring and risk assessment. In Section \ref{sec:results}, we present the results of our approach, highlighting its performance and implications through application in an example. Finally, Section \ref{sec:discussion} concludes with a discussion of the findings, their broader implications, and potential directions for future research.

\section{Motivating Application: Atopic Dermatitis} \label{sec:motivating_app}

The motivating example is a hypothetical repetition of a study investigating immunomodulatory treatments for dermatological conditions, specifically atopic dermatitis and alopecia areata (NCT03575871 in ClinicalTrials.gov). 
The external data are earlier trials studying the same conditions, and registered 
on ClinicalTrials.gov under
identifiers NCT02780167, NCT03349060, NCT03715829, and
NCT03732807.
These studies focus on evaluating the safety and efficacy of
investigational drugs. Atopic dermatitis, a chronic inflammatory skin
condition, affects approximately 15--20\% of children and 1--10\% of
adults globally, while alopecia areata, an autoimmune disorder causing
hair loss, has a lifetime prevalence of ~2\%. The conditions
represent significant public health challenges, necessitating robust
clinical research to develop effective therapies.

The selected  trials share a high degree of consistency in design and
execution. All are randomized, placebo-controlled studies involving
adolescent and adult populations (aged $\geq$12 years), and they
evaluate similar classes of oral immunomodulatory agents. The
endpoints across studies include common dermatological outcomes such
as symptom improvement in atopic dermatitis (e.g., NCT02780167,
NCT03575871) and hair regrowth in alopecia areata (e.g., NCT03349060,
NCT03732807).
These shared characteristics, in study structure, target
populations, interventions, and adverse event (AE) monitoring, make the
trials a suitable and coherent dataset for the proposed method, which
enables borrowing of information across studies to compare treatment
effects, assess heterogeneity, and estimate AE incidence rates in the
proposed Bayesian nonparametric framework.

\section{Inference Model}\label{sec:inference_model}

\subsection{Notation and Data Structure} 
We begin with an overview of the notations used to define studies, covariates, experimental units at various levels of aggregation,  clustering of units, and cluster-specific incidence rates. 
Let $j = 1, \ldots, J$ index available studies, 
including the current study and historical studies. The current study (\( j = 1 \)) is the primary focus and typically  includes 
patient-level data. 
For other studies \((j = 2, \ldots, J)\), data may be available only at different  resolutions:  
\textbf{(1)} patient-level  data  (\( j = 2, \dots, J_1 \)), which  in particular allow AE summaries reported by unique   joint covariate levels; 
\textbf{(2)}  studies providing summaries of AE outcomes stratified (marginally) by individual covariate levels 
(\( j = J_1 + 1, \dots, J_2 \));
\textbf{(3)} studies with only study-level summaries (\( j = J_2 + 1, \dots, J \)), such as aggregated counts and summary statistics for covariates.  We refer to (1), (2) and (3) as studies with \textbf{patient-level}, \textbf{covariate-summary} and \textbf{study-level} types of AE data.

Let \( d \in \mD \equiv \{ 1, \dots, D\} \) index  patient covariates. 
These may include continuous variables (e.g., age), binary indicators (e.g., treatment status),  discrete categorical,  or composite measures (e.g., multi-category conditions). For a specific discrete covariate $d$, let $e = 1, \ldots, E_d$ denote its levels, where $E_d$ is the total number of the levels. For continuous covariates, $E_d = 1$. 
\textbf{Cohorts} are subsets of patients within studies, defined based on the granularity of available data. Let $L_j$ denote the number of cohorts within study $j$. For \textbf{patient-level studies}, cohorts correspond to all unique combinations of covariates, with $L_j = \prod_{d \in \mD} E_d$; for \textbf{covariate-summary studies}, cohorts reflect the combinations of available covariates, with $L_j = \prod_{d \in \mD'} E_d$, where $\mD' \subset \mD$; for \textbf{study-level data}, the entire study constitutes a single cohort $L_j = 1$. This structure ensures that cohorts are aligned with the level of detail in each study type.

The fundamental experimental units for the upcoming model construction are \textbf{units}  ($u = 1, \ldots, U$),  representing individual cohorts within studies. A unit ($u \equiv (j, \ell)$) is indexed by its study ($j$) and cohort ($\ell = 1, \ldots, L_j$).
The total number of units is $U = \sum_{j=1}^J L_j$. For each unit $u$, let $\bx_u = (x_{u1}, \ldots, x_{uD})$ denote its $1\times D$ covariate vector. The covariate information of all the units is then represented by $\bX = \{\bx_1, \ldots, \bx_U\}$.
Both cohorts and units refer to subsets of patients within studies in this manuscript. We use the terms interchangeably: “cohorts” aligns with familiar terminology in clinical research, while “units” provides a convenient framework for model formulation and indexing.

 Next, we need notation for the AE outcomes.  We assume binary outcomes for AEs, which are reported as counts for corresponding patient populations within units. Let  \(\by = \{y_1, \ldots, y_U\}\) denote the observed AE counts across \( U \) units, and  \( \bt = \{t_1, \ldots, t_U\} \)
represent the total exposure time  (or other relevant measure of exposure)  for these units. These counts form the basis for estimating AE incidence rates while accounting for exposure.

Finally, we define some notation related to arranging units in clusters (of units). Importantly, cluster arrangements are treated as an unknown parameter, i.e., as a random quantity. 
To capture patterns and dependencies in the data, we cluster the \( U \) units into \( K \) nonempty and exhaustive subsets. Let  $\{1,\ldots, U\} = \bigcup_{k=1}^K S_k$ denote a partition \(\rho = \{S_1, \ldots, S_K\}\) of units. We will construct the partition such that  each subset \( S_k \) contains units that are considered similar based on covariate information. The partitioning divides \( \{1, \ldots, U\} \) into disjoint subsets \( S_k \), allowing us to pool information across similar units.

For each cluster \( k \), we define a cluster-specific incidence rate \( \theta_k^\star \)  as the average  rate of AEs within that cluster.  Let  \(\bth^\star = \{\theta_1^\star, \ldots, \theta_K^\star\}.\)  Throughout, we use subscript $\star$ to signify cluster-specific quantities. 
These rates \( \theta_k^\star \) are estimated using the AE counts and exposure times within each cluster, leveraging the similarity of units in the same cluster to improve precision  and marginalizing w.r.t. the random cluster arrangement $\rho$.  We define $\theta_u$ as the incidence rate for unit $u$.  Let $\theta_u = \theta_k^\star$ for units $u$ in subset $k$. We will report inference on $\theta_u$ marginalizing w.r.t. the cluster arrangement $\rho$.

The random partition $\rho$, together with the cluster-specific incidence rates $\ths_k$, implements regression of AE rates on cohort-specific covariates as
\(p(\by \mid \bx) = \int p(\by \mid \rho, \bth^\star) d p(\bth^\star) d p(\rho \mid \bx)\)
(see later for details on the assumed probability models). The marginalization is carried out w.r.t. the posterior distribution on $\rho$ and $\bth^\star$, and always conditional on cohort-specific covariates. One important feature of setting up the regression by way of a random partition is that the partition remains well defined also with varying dimension covariate vectors. 
We shall introduce a random partition model that makes use of available covariates, but does not require imputing or adjusting for not reported covariates. This is important in the context of AE monitoring, as not all studies and cohorts record the same sets of covariates.

\subsection{Clustering}
The partition $\rho$ is treated as a random variable, reflecting
uncertainty in how units should be grouped.
 We start the model construction with a prior on $\rho$.
We use an instance of a 
product partition model  (PPM)
\citep{barry1993bayesian, crowley1997product}. A PPM is
 written as 
\begin{equation}\label{eq:PPM}
    p(\rho) \propto \prod_{k=1}^K c(S_k).
\end{equation}
where $c(S_k)$ is the cohesion function. A popular choice, \(c(S_k) =
M (|S_k| - 1)!\) with $|S_k|$ denoting the cardinality of the set
$S_k$, is known as the Chinese restaurant process. It arises as the
random partition implied by the ties of a random sample from a
Dirichlet process random measure \citep{ferguson1973bayesian,
quintana2003bayesian}.  The parameter $M$ controls the expected number
of clusters.  Large $M$ encourages more clusters on average. 
See \citet{Blasi2015tpmasi} for an
extensive discussion of a Dirichlet process random partition  and
its limitations. 

However, the PPM \eqref{eq:PPM} does not explicitly incorporate covariate information, effectively treating all units as equally likely to cluster together. To address this limitation, the PPMx \citep{muller2011product} extends the PPM by introducing a similarity measure based on observed covariates.  The similarity function is used to modify the PPM to favor co-clustering of  units with similar covariate profiles,  similar to the notion of a purity function in algorithmic hierarchical clustering. 
In our framework, we  use the PPMx prior to group units by their covariate profiles. Defining clusters of similar units will later in the model construction allow us to introduce cluster-specific AE incidence rates. The PPMx prior for this partition takes the form  
\begin{equation}\label{eq:ppmx_prior}
    p(\rho) \propto \prod_{k=1}^K c(S_k) g(\bX_k^{\star}),
\end{equation}
where $\bX_k^{\star} = \{\bx_u : u \in S_k\}$ gathers the covariates of all units in cluster $k$. The similarity function $g(\bX_k^{\star})$ measures the homogeneity of units within each cluster based on their covariate profiles, thereby guiding the partition process to reflect observed similarities in the data.  When we want to highlight the nature of \eqref{eq:ppmx_prior} as a covariate dependent random partition we also write $p(\rho \mid \bx)$, by a slight abuse of notation ($\bx$ is not a r.v.).

Here we introduce a similarity function constructed on pairwise comparisons. The motivation is that in many applications, such as the current problem, pairwise comparisons are easier to quantify than for entire sets.
The similarity function $g(\bX_k^{\star})$ is  constructed based on pairwise comparisons, 
\begin{equation} \label{eq:similarity}
    g(\bX_k^{\star}) = \frac{1}{n_k(n_k - 1)/2} \sum_{u < u' \in S_k} s(\bx_u, \bx_{u'} ),
\end{equation}
where $n_k$ denotes the number of units in cluster $k$.
The pairwise similarity score $s(\bx_u, \bx_{u'} )$ evaluates  the match of  covariates of two units $u$ and $u'$.


Using $\bxi = \{\xi_1, \ldots, \xi_D\}$ to define relative weights of covariates,  the pairwise similarity score can be defined as 
\begin{equation} \label{eq:pairwise_similarity}
    s(\bx_u, \bx_{u'} ) = \frac{1}{\Xi_{uu'}} \sum_{d = 1}^D \xi_d \cdot s_d(x_{ud}, x_{u'd}),
\end{equation}
where $\Xi_{uu'} = \sum_{d  = 1}^D \xi_d$ normalizes the contributions of covariates. The term $s_d(x_{ud}, x_{u'd})$ quantifies the similarity with respect to  each covariate $d$.  
In the application, we determine the covariate weights $\bxi = \{\xi_1, \ldots, \xi_D\}$ based on domain knowledge regarding their relative importance in predicting AEs. Specifically, we first identify the most influential covariate and assign it a weight of $\xi_d = 10$, then assign smaller weights to the remaining covariates in proportion to their estimated impact on AEs. While this choice is heuristic, we find the results to be reasonably robust to moderate variations in $\bxi$, as shown in the sensitivity analysis included in Section \ref{sec:sim}. 

The specific  choice of  $s_d(x_{ud}, x_{u'd})$ depends on the type of covariate, such as binary, continuous, or composite, and ensures flexibility in modeling diverse datasets. In some problems similarity of covariates might have to be judged for pairs or groups of variables - for example multiple biomarkers with important interactions. In that case we either evaluate one term of \eqref{eq:similarity} for the group, or - equivalently - code a new derived covariate.

To handle missing covariates,
let $O_u$ be the set of covariate indices actually observed for unit $u$, and denote $\bx_u^o = \{x_{ud} : d \in O_u\}$ the observed values.
For units assigned to cluster $k$, the observed covariate values are $\bX_k^{\star o} =\{\bx_u^o : u \in S_k\}$, and for the entire dataset of covariates, $\bX^o = \{\bx_1^o, \ldots, \bx_U^o\}$. In this case, the similarity function \eqref{eq:similarity} is modified as 
\begin{equation*}
    \tg(\bX_k^{\star o}) = \frac{1}{n_k(n_k-1)/2} \sum_{u < u' \in S_k} \ts(\bx_u^o, \bx_{u'}^o ),
\end{equation*}
and
\begin{equation*}
    \ts(\bx_u^o, \bx_{u'}^o ) = \frac{1}{\tXi_{uu'}} \sum_{d \in O_u \bigcap O_{u'}} \xi_d \cdot s_d(x_{ud}, x_{u'd}),
\end{equation*}
with $\tXi_{uu'} = \sum_{d \in O_u \bigcap O_{u'}} \xi_d$.
This approach preserves the integrity of the clustering process by retaining comparability even in the presence of missing data.

The specific form of $s_d$ is tailored to each covariate type, as detailed in Appendix \ref{sec:spe_form_score}.
By flexibly accounting for different covariate types, these similarity functions provide a nuanced way to gauge how closely two units resemble each other with respect to all reported covariates.  Any other variations to accommodate the nature of specific covariates in an application are easily accommodated. 

\subsubsection{Similarity of Intervention}
Importantly,  the PPMx model  naturally supports a seamless transition
between blinding and unblinding during safety monitoring.
 We consider treatment assignment as an additional covariate $x_{ud}$,
which is coded differently before and after unblinding for the current
study and for units from historical trials.

Upon \textbf{unblinding}, the
current study is divided into separate units based on the intervention
arms with corresponding $x_{ud}$.
If needed the intervention covariate is coded as two
covariates, including a binary covariate for
drug identity and an ordinal covariate for dose level. 
Specifically, under unblinding, the intervention covariate for unit $u$ is denoted as
$$x_{ud} \in \mathcal{T} = \{0\} \bigcup \{(g,h):g \in \mathcal{G}, h \in \mathcal{H}_g\}.$$
where $x_{ud} = 0$ indicates placebo, and $x_{ud} = (g,h)$ represents
drug identity $g \in \mathcal{G}$ and dose level $h \in \mathcal{H}_g = \{1, \ldots, H_g\}$,  including dose level $h$ if needed. 
Based on this representation, the similarity of the intervention
covariates   $x_{ud}$ and $x_{u'd}$ for two units $u$ and $u'$ is
 defined as
\[
s_d(x_{ud}, x_{u'd}) =
\begin{cases}
1, & \text{if } x_{ud} = x_{u'd} = 0 \\
1 - \dfrac{|h_{u} - h_{u'}|}{H_g}, & \text{if } x_{ud} = (g,h_{u}),\ x_{u'd} = (g, h_{u'}) \\
0, & \text{if } x_{ud} = 0 \text{ and } x_{u'd} \neq 0,\ \text{or if } g_{u} \neq g_{u'}
\end{cases}.
\]

Under \textbf{blinding} the intervention covariate for the current study
is coded differently. It is
coded to represent a mixture of control and treatment
arms without explicit differentiation. This approach preserves the
blinding while allowing the model to account for the uncertainty in
intervention allocation.
Specifically, under blinding we code $x_{ud}$ as a mixture,
represented as a finite discrete distribution, 
\( x_{ud} \sim \nu_u \), where \(
\nu_u \) is supported on \( \mathcal{T} \) and satisfies \( \sum_{r
  \in \mT} \nu_u(r) = 1 \) and \(\nu_u(r) \geq 0\) for all $r \in
\mT$. The similarity of the intervention covariates of two units \( u
\) and \( u' \) under blinding is defined as: 
\[
s_d^{b}(x_{ud}, x_{u'd}) =
\begin{cases}
s_d(x_{ud}, x_{u'd}), & \text{if both } x_{ud} \text{ and } x_{u'd} \text{ are known} \\[8pt]
\displaystyle \sum_{r \in \mT} \nu_u(r) \cdot s_d(r, x_{u'd}), & \text{if } x_{ud} \sim \nu_u,\ x_{u'd} \text{ known} \\[8pt]
\displaystyle \sum_{r \in \mT} \nu_{u'}(r) \cdot s_d(x_{ud}, r), & \text{if } x_{ud} \text{ known},\ x_{u'd} \sim \nu_{u'} \\[8pt]
\displaystyle \sum_{r_1 \in \mT} \sum_{r_2 \in \mT} \nu_u(r_1)\, \nu_{u'}(r_2) \cdot s_d(r_1, r_2), & \text{if both } x_{ud} \sim \nu_u,\ x_{u'd} \sim \nu_{u'}
\end{cases}.
\]

\subsection{Joint Distribution}



 The prior model $p(\rho)$ in \eqref{eq:ppmx_prior} is now augmented to a regression model by adding cluster specific parameters for a sampling model
\begin{equation}\label{eq:sampling_model}
    p(\by \mid \rho, \bth^\star) = \prod_k \prod_{u \in S_k} p(y_u \mid \theta_k^\star).
\end{equation}
In our case $y_u$ are the AE counts in unit $u$. The model is completed with priors for $\bth^\star$ and $\rho$. We assume gamma priors for $\ths_k$. Averaging \eqref{eq:sampling_model} w.r.t the random partition and marginalizing w.r.t $\bth^\star$ implies the desired regression. 

For later reference we state the complete model:
\begin{equation} \label{eq:joint_model}
    \begin{aligned}
    p(\by, \rho, \bth^{\star}, a, b) = & \ p(\by \mid \rho, \bth^{\star}) p(\bth^{\star} \mid \rho, a, b) p(\rho) p(a) p(b) \\
    \propto & \ \prod_{k = 1}^K \left[ \left( \prod_{u \in S_k} p(y_u \mid \ths_k) \right) p(\ths_k \mid a, b) \right] \cdot p(\rho) \cdot p(a) p(b),
\end{aligned}
\end{equation}
where $p(y_u \mid \ths_k)$ denotes the probability mass function of the Poisson distribution $\tP(t_u\ths_k)$ evaluated at the point $y_u$. 
The likelihood factor $p(\by \mid \rho, \bth^{\star})$ models the observed adverse event counts, which are assumed conditionally independent given the partition $\rho$, cluster-specific parameters $\bth^{\star}$, and the total exposure time of all units $\bt$.  The partition prior $p(\rho)$ incorporates covariate information through the similarity function $\tg(\cdot)$,
\[
p(\rho)  \propto \prod_{k = 1}^K c(S_k ) \tg(\bX_k^{\star o} ),
\]
where $\bX_k^{\star o}  = \{\bx_u^o : u \in S_k\}$ denotes the observed covariates of all units in cluster $k$. The term $\tg(\cdot)$ ensures that the clustering process is informed by observed covariate similarities.

To facilitate the explanation of the decision problem in the subsequent section and posterior inference, we introduce  the incidence rate of unit $u$, denoted as $\theta_u$,  along with the latent cluster membership indicator $c_u$, which represents the cluster assignment for unit $u$.
Given the latent cluster membership indicators,  $\theta_u = \ths_k$ when $c_u = k$. Consequently,  the likelihood in \eqref{eq:joint_model} can be expressed for each unit $u$ as
\[
y_u \mid \bth^{\star}, c_u = k  \sim \tP( t_u \ths_k).
\]
Posterior inference is performed using Markov chain Monte Carlo (MCMC) methods. Additional computational details can be found in Appendix \ref{sec:pos_computation}.

\section{Decision Problem}
\label{sec:decision_problem}
The inference model \eqref{eq:joint_model} implements the desired  multi-resolution  
regression, including borrowing of strength across units with similar covariates, and using only available covariates for each units. Posterior inference allows, for example, estimation of incidence rates in each unit. However, estimating parameters is only half the solution to the inference problem. Making (or at least recommending) decisions is the other half. In the context of aggregated AE monitoring the latter is actually the main inference goal. The aim is to  assess the safety profile of the current study by comparing it to relevant external data,  and propose  informed decisions regarding study unblinding and  mandatory reporting of  potential safety concerns.

%

%

\subsection{Decision Boundaries} \label{sec:dec_bound}

\begin{figure}[!htb]
    \centering
    \includegraphics[width=\linewidth]{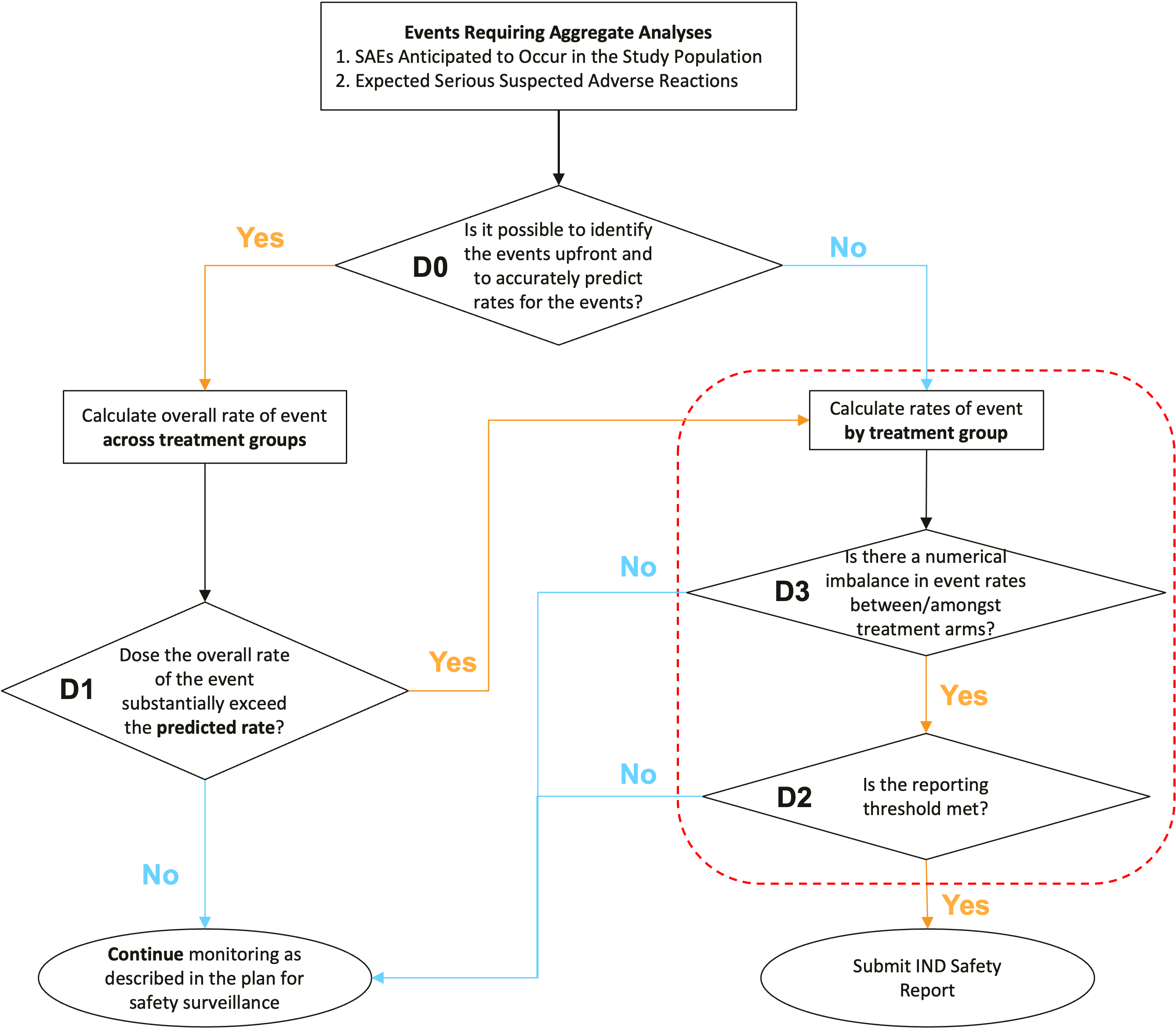}
    \caption{Aggregate Safety Analysis Process. Dashed box: process under unblinded safety review.  The decision notes (D0, D1, D2, and D3) are numbered for later reference.}
    \label{fig:flowchart}
\end{figure}

We formalize the proposed decisions—specifically, the decision nodes in Figure~\ref{fig:flowchart} except D0, which lies beyond the scope of statistical expertise and relies primarily on clinical or operational judgment-by defining  decision boundaries on  posterior probabilities for clinically relevant events, that is, events in terms of the unknown AE rates. We use three different boundaries to recommend unblinding, to judge imbalance of treatment versus control arm in the current trial and to recommend reporting.

As described before, let \( j = 1 \) and \( u = 1, \ldots, L_1 \) denote the current study and its associated units under blinding, respectively. Define \( \mU = \{1, \ldots, U\} \) as the set of indices for all units and \( \mU_1 = \{1, \ldots, L_1\} \) as the subset of indices corresponding to the units in the current study.  Naturally, under blinding the cohorts of the current study in $\mU_1$ do not include stratification by treatment. 

Under \textbf{blinding}, we monitor the safety profile of the current study by comparing the average incidence rate  within \( \mU_1 \) to a weighted  average of  incidence rates for units outside the current study \( \mU \setminus \mU_1 \).  
We use weights based on pairwise similarity of pairs of units $u$ and $u'$, $\ts(\bx_u,\bx_{u'} )$. For simplicity, we write $\ts_{uu'}$. For each $u \in \mU_1$ and $u' \in \mU \setminus \mU_1 $, we define
\[
w_{uu'} = \frac{\ts_{uu'}}{\sum_{u' \in \mU \setminus \mU_1} \ts_{uu'}}.
\]
Let
\begin{equation}\label{eq:E1}
    E_1 \equiv \left\{\frac{1}{L_1} \sum_{u \in \mU_1} \theta_u > \frac{1}{L_1} \sum_{u \in \mU_1}\sum_{u' \in \mU \setminus \mU_1} w_{uu'}\theta_{u'} + \delta\right\},
\end{equation}
and 
\begin{equation} \label{eq:pi_blind}
    \pi_1 \equiv \Pr\left( E_1 \mid \by\right),
\end{equation}
where $\delta$ is a clinically minimum meaningful difference. The probability in \eqref{eq:pi_blind} is marginalized w.r.t. all unknown parameters, including in particular the random partition of units and the cluster specific parameters. We recommend unblinding if $\pi_1 > \lambda_1$, where \( \lambda_1 \) is a predefined decision boundary indicating when the observed incidence rate in the current study is sufficiently higher than the weighted incidence rate outside the study to warrant unblinding.  This decision boundary implements decision node D1 in Figure \ref{fig:flowchart}.

Upon \textbf{unblinding}, the current study is divided into distinct units according to intervention arms, allowing us to distinguish the treatment arm from other arms more precisely. Let  \( \wmU \) denote the set of indices for all units following unblinding. Let \( \wmU_{1T}\) be the subset corresponding to the treatment-arm units in the current study, and $\wL_{1T} = |\wmU_{1T}|$. Let
\begin{equation}\label{eq:E2}
    E_2 \equiv \left\{ \frac{1}{\wL_{1T}} \sum_{u \in \wmU_{1T}} \theta_u > \frac{1}{\wL_{1T}} \sum_{u \in \wmU_{1T}}\sum_{u' \in \wmU \setminus \wmU_{1T}} w_{uu'}\theta_{u'} + \delta \right\},
\end{equation}
and 
\begin{equation} \label{eq:pi_unblind}
    \pi_2 \equiv \Pr\left( E_2 \mid \by\right),
\end{equation}
where the weights in \eqref{eq:pi_unblind} are computed as $w_{uu'} = \ts_{uu'} / \sum_{u' \in \wmU \setminus \wmU_{1T}} \ts_{uu'}$ for $u \in \wmU_{1T}$ and $u' \in \wmU \setminus \wmU_{1T} $.
We then assess toxicity in the treatment arm by  comparing with another threshold,  $\pi_2 > \lambda_2$.
If this condition is satisfied, the treatment arm is deemed excessively toxic, thus necessitating the submission of an IND safety report.  This decision boundary implements a representative criterion at decision node D2 in Figure~\ref{fig:flowchart}, selected from among the various options discussed in \citet{us2021sponsor}. 

Following the FDA guidance \citep{us2021sponsor}, under the proposed model we can evaluate any additional reporting thresholds. For example, we may also compare the treatment-arm unit(s) with the control-arm unit(s) in the current study after unblinding. 
Let
\begin{equation}\label{eq:E3}
    E_3 \equiv \left\{ \frac{1}{\wL_{1T}} \sum_{u \in \wmU_{1T}} \theta_u > \frac{1}{\wL_{1C}} \sum_{u \in \wmU_{1C}}\theta_{u} + \delta \right\},
\end{equation}
and 
\begin{equation} \label{eq:pi_trt_ctrl}
    \pi_3 \equiv \Pr\left( E_3 \mid \by\right),
\end{equation}
where \( \wmU_{1C}\) represents the subset of the control-arm units in the current study with cardinality $\wL_{1C}$.
The reporting threshold is formalized as $\pi_3 > \lambda_3$.
If this condition holds, the treatment arm is deemed excessively toxic compared to the control arm,  necessitating the submission of an IND safety report.  This decision boundary implements decision node D3 in Figure \ref{fig:flowchart}. 

In summary, we propose to impose a solution to the decision problem by establishing decision boundaries for \eqref{eq:pi_blind},  \eqref{eq:pi_unblind}, and \eqref{eq:pi_trt_ctrl}. We propose to set the thresholds $\lambda_1$, $\lambda_2$ and $\lambda_3$ based on considering frequentist operating characteristics under some plausible scenarios, including type I error probabilities and power over repeat simulations. 
Here a type I error rate is defined as the probability (over repeat simulations) of recommending unblinding by $\pi_1 > \lambda_1$ and of recommending reporting by $\pi_2 > \lambda_2$ and $\pi_3 > \lambda_3$ when $E_1$, $E_2$ and $E_3$ are all false under the simulation truth; and power refers to the probability (again, under repeat simulation) of (correctly) recommending unblinding and reporting, respectively, when $E_1$, $E_2$ and $E_3$ are all true under the simulation truth. 
Considering additional reporting thresholds, we may  replace $E_2$ with $E_3$ or $E_2 \bigcup E_3$, depending on investigators' need in practice. 


\subsection{Pre-trial Background Rate} \label{sec:pre_br}
In addition to the reporting requirements  the FDA guidance \citep{us2021sponsor} also requires that a background rate (i.e., the predicted rate of decision node D1 in Figure \ref{fig:flowchart}) be already stated in the investigator brochure as a reference before the initiation of the current study. At this stage, no patients have been enrolled, and no data from the current study are available. 
Therefore, instead of using covariates from collected data, we may consider using covariates associated with the eligibility criteria. Further details are provided in Appendix \ref{sec:supp_pre_br}.

\section{ Case Study  } 
\label{sec:results}

\begin{figure}[!htb]
    \centering
    \begin{subfigure}{.49\textwidth}
    \centering
    \includegraphics[width=\linewidth]{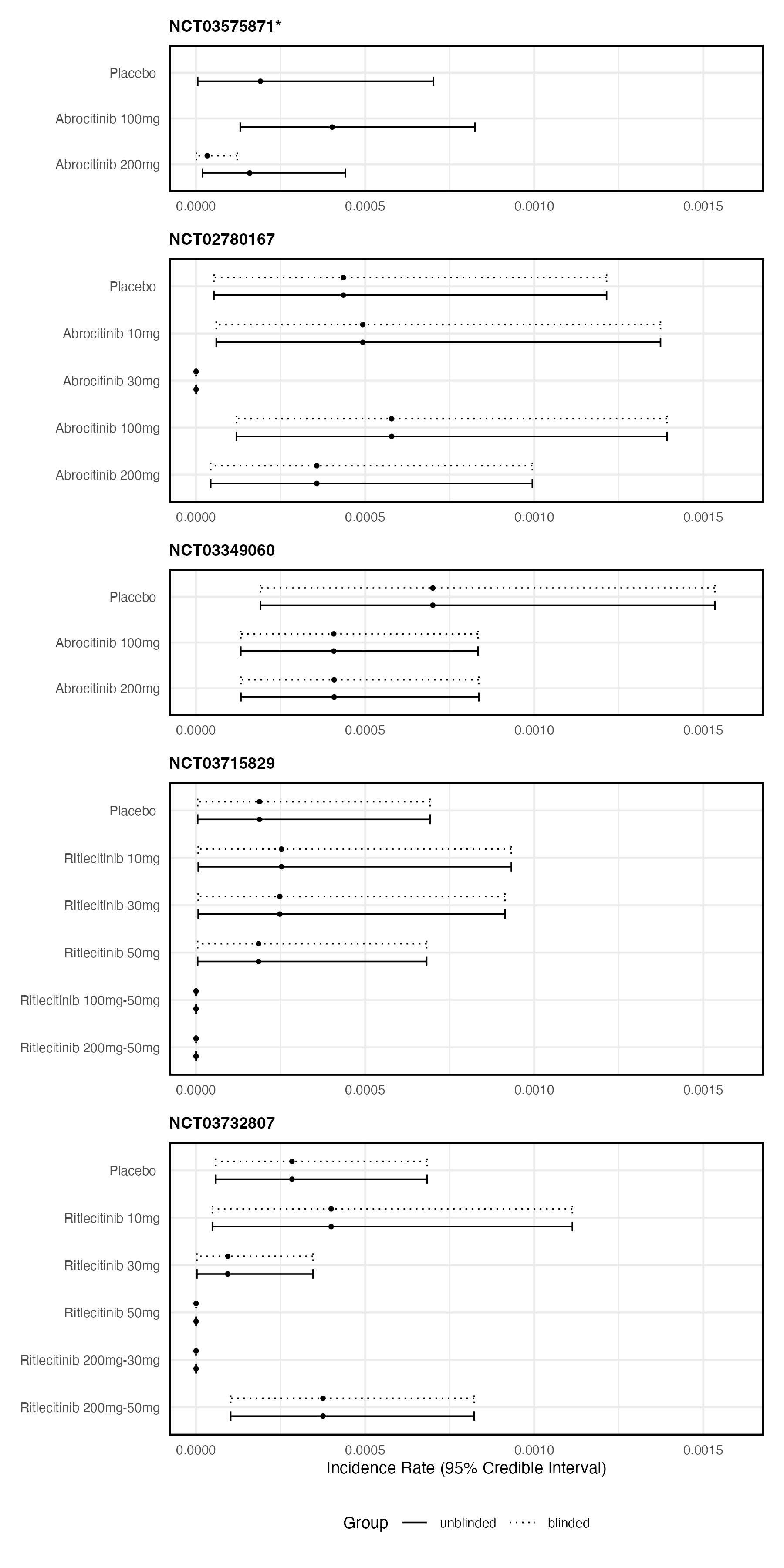}
    \caption{Independent Poisson Model}\label{fig:analysis_ind}
    \end{subfigure}
    \begin{subfigure}{.49\textwidth}
    \centering
    \includegraphics[width=\linewidth]{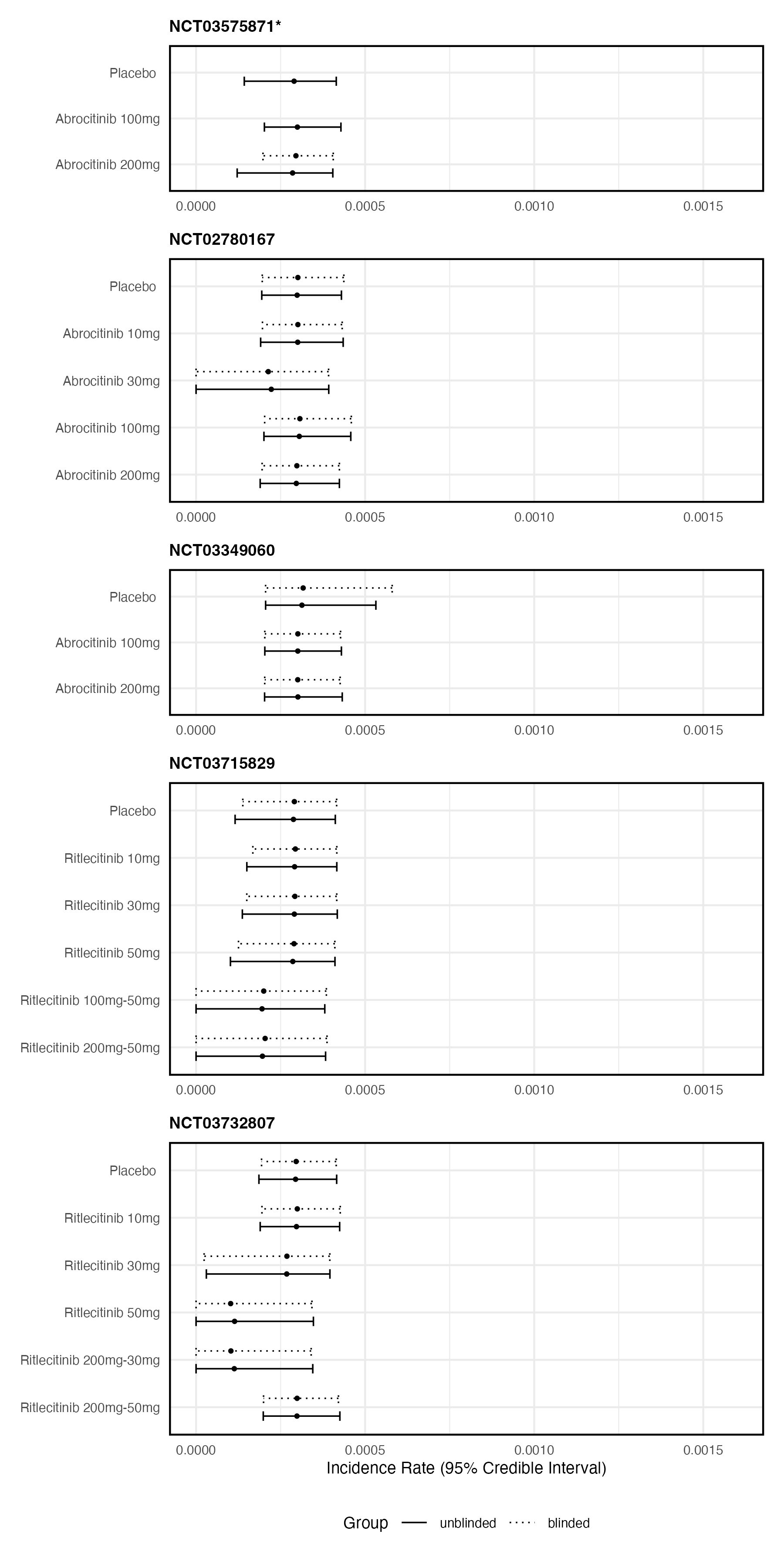} 
    \caption{PPMx}\label{fig:analysis_PPMx}
    \end{subfigure}
    \caption{Forest plot of posterior mean AE incidence rates by unit, with 95\% credible intervals, for the atopic dermatitis case study described in Section \ref{sec:motivating_app}. (a) under an independent Poisson model $y \sim \text{Poi}(t\theta)$ with an improper prior $\theta \propto 1$, fitted separately for each unit. (b) under the proposed PPMx.}
    \label{fig:analysis_res1}
\end{figure}

We reanalyze data for the studies described in Section \ref{sec:motivating_app}, treating NCT03575871 as the current trial. 
When the current study is still blinded we observe 21 units (cohorts); after unblinding, stratification by the intervention increases the total to 23 units.
Table~\ref{tab:data_set1} summarizes all contributing studies, their covariates, observed numbers of SAEs and empirical incidence rates (IRs).
 We used five covariates to define the similarity scores, 
their types and relative weights   listed in  Table~\ref{tab:cov_xi}. 
The  total mass  $M$ of $c(S_k )$ in \eqref{eq:ppmx_prior} is set to 2. A $\text{Gamma}(1,1)$ hyperprior is placed on both hyperparameters $a$ and $b$ in $\theta_k \sim \text{Gamma}(a, b)$.  Figure~\ref{fig:analysis_res1} shows a forest plot of posterior IR estimates. 

Figure~\ref{fig:analysis_res1} displays the unit–specific posterior IR estimates obtained (a) from an independent Poisson model, $y_u \sim \text{Poi}(t_u\theta_u)$ with the improper prior $p(\theta_u) \propto 1$, and (b) under the proposed PPMx model. Several qualitative differences stand out. Figure~\ref{fig:analysis_ind} shows substantial between-unit variability in interval length: units with low exposure have very wide 95\% intervals, in several cases spanning more than a two-fold change on the rate scale. By contrast, in Figure~\ref{fig:analysis_PPMx} virtually all intervals are shorter  reflecting how  the PPMx model pulls information from covariate-matched units. The precision gain is most pronounced for low-exposure historical cohorts and for small dose-escalation arms in the current study.

Under the independent analysis every unit estimate is free to drift. Consequently Figure~\ref{fig:analysis_ind} displays several extreme point estimates—particularly for the unit without any adverse events—that appear disconnected from the overall pattern of the data. In Figure~\ref{fig:analysis_PPMx} those same units move appreciably toward the overall pattern of their covariate-defined neighbors. This adaptive shrinkage stabilizes the estimated safety profile without masking real signals: units that remain high after shrinkage are now supported by more coherent evidence.

In summary, the independent Poisson model provides a raw picture of the data but delivers uneven and often inadequate precision, especially for rare-event units. The PPMx analysis respects the heterogeneity detailed in Table \ref{tab:data_set1} through guided clustering, produces more stable, interpretable IR estimates that are better suited for the safety-signal decision criteria developed in Section \ref{sec:decision_problem}.

\section{Simulation}\label{sec:sim}
We carried out simulations to evaluate the operating characteristics of the proposed PPMx approach based on 1,000 independent repeat simulations. Results are compared with the methods proposed by \citet{cai2010meta} and \citet{mukhopadhyay2018bayesian},  under three scenarios. 
The first  is a null scenario (“scenario 0”), assuming no true excess AE rates in the current study. 
The second (“scenario 1”) assumes true AE rates beyond the thresholds for  $E_1$ and $E_2$, but not $E_3$, reflecting a partially positive case. 
The third (“scenario 2”) assumes  elevated true AE rates  that should trigger all three events: $E_1$, $E_2$, and $E_3$. The specific simulation scenarios are summarized in Table \ref{tab:sim_sc}. For simplicity, we adopt the same units, covariates, and total exposure times as those in Table \ref{tab:data_set1}, and vary only the assumed hypothetical IRs for each unit across scenarios. These hypothetical IRs are assumed to depend solely on the intervention and its dose.

For each simulation, the number of AEs per unit is generated from a Poisson distribution, with the mean equal to the product of the unit specific exposure time and its scenario-specific hypothetical IR. The  total mass  $M$ of $c(S_k )$ in \eqref{eq:ppmx_prior} is set to 2. We place a $\text{Gamma}(1,1)$ hyperprior on both hyperparameters $a$ and $b$ in $\ths_k \sim \text{Gamma}(a,b)$. 

Simulation results are presented in Table \ref{tab:sim_res}. \citet{mukhopadhyay2018bayesian}, labeled as ``MUK", considers only the unblinding decision. Therefore, we compare only the frequency of simulations in which  unblinding is recommended. Under the null scenario, both PPMx and \citet{mukhopadhyay2018bayesian} control the frequency of false positives at 5\%, through calibration of the decision threshold. In this setting, PPMx and \citet{mukhopadhyay2018bayesian} exhibit similar performance in scenarios 1 and 2.

In contrast, \citet{cai2010meta}, labeled as ``CAI", considers only 
the decisions of ``unblinding and reporting" $E_1 \bigcap (E_2 \bigcup E_3)$ (without distinguishing $E_2$ versus $E_3$).
Accordingly, we compare the frequency of simulations in which $E_1 \bigcap (E_2 \bigcup E_3)$ is reported.
Here, $E_1 \bigcap (E_2 \bigcup E_3)$ refers to simulations where $E_1$ and at least one of $E_2$ or $E_3$ are declared as true signals.
In scenario 1, the frequency of $E_1 \bigcap (E_2 \bigcup E_3)$ is controlled at 20\% through calibration of the decision thresholds  to match the corresponding rate under the proposed scheme and allow comparison.  Under this setting, we find that PPMx and \citet{cai2010meta} exhibit better performance in scenario 2.

For sensitivity analysis, we include ``PPMx-L" as an alternative method, which applies the same  proposed  PPMx framework but with fewer covariates. Specifically, we assume that the covariates stgAge and Sex are unavailable for all studies and conduct simulations under this reduced-covariate setting. As shown in the following table, PPMx-L still demonstrates performance comparable to that of the full PPMx model.

\section{Discussion}
\label{sec:discussion}

The proposed PPMx framework offers  a model-based and principled approach to using real world data (RWD)  for 
monitoring  aggregate AEs.  
Unlike previous methods that apply uniform borrowing across trials, PPMx incorporates known trial-specific features through a covariate-dependent partitioning strategy, enhancing interpretability and relevance. By leveraging pairwise similarity measures and cluster-level borrowing, PPMx enables more precise comparisons across studies or subpopulations. 
 Casting unblinding and the comparisons of AE rates as actions in a decision problem allows careful and systematic consideration of the proposed recommendations. 

 Some limitations remain. 
First, the inference model currently treats each AE independently, without accounting for the potential dependence structure across multiple, jointly occurring AEs. This simplification may overlook informative correlations among related safety signals.  
Second, the framework assumes that the relationships among available covariates are not confounded by unmeasured risk factors, which may not always hold in practice. In particular, two or more covariates may share dependence through an unobserved confounder, potentially biasing the clustering structure. 
Sensitivity analyses or adjustments for potential confounding could improve robustness in such settings.

Several directions can be explored to extend the proposed PPMx framework. One natural extension is to treat each patient as the experimental unit  in cases where patient-level data were available.  This would enable even finer-grained clustering and borrowing, potentially enhancing the model’s resolution in heterogeneous populations. Another promising direction is to incorporate prior distributions on the similarity weights, allowing more flexible modeling of the relative importance of different covariates in the partitioning process.  Lastly, the AE events $E_1$, $E_2$, and $E_3$ are based on mean AE rates. One could use more robust summary such as median or quantiles in defining these events.  

In addition to methodological extensions  in the inference model,  future work could aim to incorporate a broader set of safety signal detection criteria in alignment with evolving FDA guidance \citep{us2021sponsor}. Some thresholds—such as predefined incidence rate increases or event clustering within specific time windows—are relatively easy to formalize within a statistical framework. For example, temporal relationships (e.g., early AE onset post drug initiation), consistency of increased AE rates across multiple trials, and patient-level patterns (e.g., higher AE rates in susceptible subgroups) can be operationalized through time-to-event models, interaction terms, or stratified analyses. However, other important considerations—such as the presence of a plausible mechanism of action, nonclinical evidence from animal or genetic studies, or pharmacological insights related to drug targets and class effects—are more difficult to formalize statistically. These components require integration of biological knowledge and expert judgment, underscoring that safety monitoring extends beyond statistical signal detection and into the realm of mechanistic plausibility and clinical interpretation. 

Thus, while our proposed PPMx framework provides a structured and interpretable tool for data-driven monitoring, it is best understood as one component within a broader, multidisciplinary safety evaluation strategy. Importantly, safety monitoring is not solely a statistical exercise but a complex process that integrates statistical modeling with clinical insight, operational logistics, and regulatory considerations. The PPMx framework serves as a foundational statistical layer within this larger context, providing principled, interpretable outputs to support timely and informed safety decisions.

\bibliographystyle{apalike}
\bibliography{AE_monitor}

\section*{Acknowledgments}
We have no conflicts of interest to
declare.
We used AI-based tools to assist with language polishing and editorial refinement of the manuscript.



\section*{Data Availability Statement}
The data  used for the application of the proposed method are available in the Supporting Information. The analysis code is also hosted on GitHub at \href{https://github.com/SJ-24/PPMx}{https://github.com/SJ-24/PPMx}.

\begin{table}[!htb]
\centering
\caption{Covariates and their types and relative weights $\xi_d$'s for the atopic dermatitis case study.}
\label{tab:cov_xi}
\begin{tabular}{|c|c|c|}
\hline
\textbf{Covariate} & \textbf{Covariate Type} & $\xi_d \in (0,10]$ \\ \hline
Intervention + Dose      &  Binary + Ordinal     & 10      \\ \hline
Condition  & Binary      & 5      \\ \hline
Phase      & Composite     & 4      \\ \hline
Study      & Binary      & 4      \\ \hline
stgAge      & Composite      & 2      \\ \hline
Sex      & Binary      & 2      \\ \hline
\end{tabular}
\end{table}

\begin{table}[!htb]
\scriptsize
  \centering
  \begin{threeparttable}
  \caption{Summary of simulation scenarios and results.  Each row represents a different simulation scenario. The columns under Average Incidence Rate show the average true IRs across units, including the current study (Curr) and background (Bg) rates under blinding, and the treatment (Trt), control (Ctl), and background (Bg) rates under unblinding. Values of average IRs are scaled by \(10^{-4}\) and rounded to two decimal places for readability.
  The columns under Operating Characteristics (OC) summarize the performance of four methods (PPMx, MUK, CAI, PPMx-L).
  Specifically, $E_1$ denotes the frequency of simulations with which $E_1$ is declared as a true signal. The expression $E_1 \bigcap (E_2 \bigcup E_3)$ represents the frequency when $E_1$ and at least one of $E_2$ or $E_3$ are declared as true signals. 
    } \label{tab:sim_res}
  \renewcommand{\arraystretch}{1.2} 
  \begin{tabular}{c|cc|ccc|cc|c|c|cc}
    \hline
    \multirow{3}{*}{\textbf{Scenario}} 
      & \multicolumn{5}{c|}{\textbf{Average Incidence Rate}} 
      & \multicolumn{6}{c}{\textbf{Operating Characteristics}} \\
    \cline{2-12}
      & \multicolumn{2}{c|}{Blinding} 
      & \multicolumn{3}{c|}{Unblinding} 
      & \multicolumn{2}{c|}{PPMx} 
      & \multicolumn{1}{c|}{MUK} 
      & \multicolumn{1}{c|}{CAI} 
      & \multicolumn{2}{c}{PPMx-L} \\
    \cline{2-12}
      & Curr & Bg 
      & Trt & Ctl & Bg 
      & $E_1$ & $E_1 \bigcap (E_2 \bigcup E_3)$ 
      & $E_1$ & $E_1 \bigcap (E_2 \bigcup E_3)$ 
      & $E_1$ & $E_1 \bigcap (E_2 \bigcup E_3)$ \\
    \hline
    0 & 3   & 3   & 3   & 3   & 3   & 5\%$^*$  & 0    & 5\%  & 0    & 5\%  & 0 \\
    1 & 5.95   & 4.22 & 5.94 & 6   & 4.48 & 56.6\% & 20\%$^{**}$ & 57.3\% & 20\% & 51.8\% & 20\% \\
    2 & 6.66 & 4.92 & 7.43 & 3   & 4.88 & 74.3\% & 59.4\% & 73.4\% & 57.3\% & 68.4\% & 57\% \\
    \hline
  \end{tabular}
  \begin{tablenotes}
		\item[*] Value fixed at 5\% by design, through calibration of $\lambda_1$ for \eqref{eq:pi_blind}. The same applies to the results reported for MUK \citep{mukhopadhyay2018bayesian} and PPMx-L.
            \item[**] Value fixed at 20\% by design, through calibration of $\lambda_2$ and $\lambda_3$ for \eqref{eq:pi_unblind} and \eqref{eq:pi_trt_ctrl}. The same applies to the results reported for CAI \citep{cai2010meta} and PPMx-L.
    \end{tablenotes}
  \end{threeparttable}
\end{table}

\begin{appendices}
\appendixpage
\setcounter{table}{0}
\renewcommand{\thetable}{A.\arabic{table}}
\setcounter{equation}{0}
\renewcommand{\theequation}{A.\arabic{equation}}
\setcounter{figure}{0}
\renewcommand{\thefigure}{A.\arabic{figure}}

\section{Specific Forms of Similarity Score} \label{sec:spe_form_score}
For \textbf{continuous covariates}, we define 
\[
s_d(x_{ud}, x_{u'd}) =  k(x_{ud} - x_{u'd}) \equiv  e^{- \frac{1}{\gamma_d^2} (x_{ud} - x_{u'd})^2},
\]
with a scale factor $\gamma_d$.
For \textbf{binary covariates},
\[
s_d(x_{ud}, x_{u'd}) =  \bone(x_{ud} = x_{u'd}),
\]
where $\bone(\cdot)$ is the indicator function.
For \textbf{categorical covariates},
\[
s_d(x_{ud}, x_{u'd}) =  |x_{ud} - x_{u'd}|,
\]
where $x_{ud}$ and $x_{u'd}$ are both proportions.
For \textbf{ordinal covariates},
\[
s_d(x_{ud}, x_{u'd}) =  1 -  \frac{|x_{ud} - x_{u'd}|}{E_d},
\]
where $E_d$ is the total number of levels for covariate $d$.
For \textbf{composite covariates},
\[
s_d(x_{ud}, x_{u'd}) = s^*(x_{ud}, x_{u'd}),
\]
where $s^*(x_{ud}, x_{u'd})$ accounts for partial matches:
\[
s^*(x_{ud}, x_{u'd}) = 
\begin{cases} 
      1, & \text{if $x_{ud}$ and $x_{u'd}$ are fully matched}, \\ 
      \frac{1}{2}, & \text{if they are partially matched}, \\
      0, & \text{if they are disjointed}.
\end{cases}
\]

\section{Posterior Computation} \label{sec:pos_computation}
Posterior inference is performed using MCMC methods, specifically employing a combination of Metropolis-Hastings (MH) and Gibbs sampling  transition probabilities.  The following updates are implemented in each iteration:

\begin{enumerate}
    \item \textbf{Update $a$}: The shape parameter $a$ is updated using  a MH transition probability  with a log-normal proposal distribution. 
    Specifically, a proposed value $a'$ is sampled from $a' \sim \text{LogNormal}(\log(a), \sigma^2_{\text{proposal}})$. The acceptance probability is calculated as  $\min{1, r}$ with 
\[
r = \frac{p(a' \mid \bth^\star, b, \alpha_a, \beta_a) q(a \mid a')}{p(a \mid \bth^\star, b, \alpha_a, \beta_a) q(a' \mid a)},
\]
where $p( \cdot \mid \bth^\star, b, \alpha_a, \beta_a)$ is the posterior density combining the gamma prior and the likelihood from $\theta_k^\star \mid \rho \sim \text{Gamma}(a, b)$, and $q(\cdot \mid \cdot)$ is the log-normal proposal density. The proposed value $a'$ is accepted with probability $\min(1, r)$; otherwise, the current value of $a$ is retained.

    \item \textbf{Update $b$}: The rate parameter $b$ is updated using Gibbs sampling from the conditional posterior distribution:
    \[
    b \sim \text{Gamma}(aK + \alpha_b, \sum_{k} \ths_k + \beta_b),
    \]
    where $K$ is the current number of clusters.

    \item \textbf{Update $\rho$}: The partition $\rho$ is updated based on Algorithm 8 from \citet{neal2000markov}, which is realized by updating the latent cluster membership indicator $c_u$ for each unit $u$. The full conditional probability of $c_u = h$, where $h = 1, \ldots, K^{-u}, K^{-u}+1, \ldots, K^{-u}+m$, is given by:
    \[
    \Pr(c_u = h \mid -) =
    \begin{cases}
      \text{Poi}(y_u; t_u \theta_h^{\star}) \frac{c(S_h^{-u} \bigcup \{u\}) \widetilde{g}(\bX_h^{\star o (-u)} \bigcup \{\bx_u^o\})}{c(S_h^{-u}) \widetilde{g}(\bX_h^{\star o (-u)})}, & h = 1, \ldots, K^{-u}, \\
      \text{Poi}(y_u; t_u \theta_{\text{new}, h}^{\star}) c(\{u\}) \widetilde{g}(\{\bx_u^o\}) / m, & h = K^{-u}+1, \ldots, K^{-u}+m,
    \end{cases}
    \]
    where all the symbols with the superscript $-u$ or $(-u)$ correspond to the partition $\rho^{-u}$  of all units excluding unit $u$:  
    \begin{itemize}
        \item $K^{-u}$: The number of clusters in $\rho^{-u}$.
        \item $S_k^{-u}$: Cluster $k$ in $\rho^{-u}$.
        \item $\bX_h^{\star o (-u)}$: The observed covariate values for units assigned to cluster $k$ in $\rho^{-u}$.
    \end{itemize}
    Here, $m$ is a prespecified number of auxiliary components, and $\theta_{\text{new}, h}^{\star}$ is is a cluster-specific incidence rate for a newly proposed cluster, sampled from a $\text{Gamma}(a, b)$ distribution.

    \item \textbf{Update $\bth^{\star}$}: For each cluster $k$, the cluster-specific incidence rate $\ths_k$ is updated using a Gibbs sampling transition probability from:
    \[
    \ths_k \sim \text{Gamma}\left(\sum_{c_u = k} y_u + a, \sum_{c_u = k} t_u + b\right).
    \]
\end{enumerate}

\section{Pre-trial background rate} \label{sec:supp_pre_br}
In addition to the reporting requirements  the FDA guidance \citep{us2021sponsor} also requires that a background rate be already stated in the investigator brochure as a reference before the initiation of the current study. At this stage, no patients have been enrolled, and no data from the current study are available. We begin by estimating the incidence rate for each unit \( u \).  Recalling  $c_u$ defines the cluster membership indicator for unit $u$  and then $\theta_u = \ths_{c_u}$, let $\widehat{\theta}_u = \E[\theta_u \mid \widetilde{\by}]$, where where $\widetilde{\by}$ is the available historical data. 
Since no data from the current trial are available at this point, we only estimate incidence rates for units in $\mU \setminus \mU_1$, where $\mU$ denotes all the units in the current study and historical studies, and $\mU_1$ the units in the current study.

Following the definition of $E_1$ in \eqref{eq:E1}, we define the background rate $\ths_0$ as
\begin{equation} \label{eq:back_undef}
    \ths_0 = \frac{1}{L_1} \sum_{u \in \mU_1}\sum_{u' \in \mU \setminus \mU_1} w_{uu'}\hth_{u'}
\end{equation}
However, the weight $w_{uu'}$ cannot be directly computed due to the  still missing  of current study data. To address this, we propose defining the weights based on covariates derived from the common eligibility criteria shared by the historical and current studies. 

For each unit $u$, let $\bx^e_u$ denote its covariate vector associated with eligibility criteria. We compute the pairwise similarity $\ts(\bx^e_u,\bx^e_{u'})$. Similar to \eqref{eq:pairwise_similarity}, $\bxi^e$ is used to emphasize or downplay specific eligibility covariates within the eligibility criteria, i.e., their relative weights. 
For simplicity, we write $\ts^e_{uu'}$. The weight is then defined as, for each $u \in \mU_1$ and $u' \in \mU \setminus \mU_1 $,
$$w^e_{uu'} = \frac{\ts^e_{uu'}}{\sum_{u' \in \mU \setminus \mU_1} \ts^e_{uu'}}.$$

By replacing $w_{uu'}$ in \eqref{eq:back_undef} with the eligibility-based similarity weight
$w^e_{uu'}$, the background rate $\ths_0$ becomes computable even before the current study begins.


\newpage
\section{Analysis Settings and Additional Results} \label{sec:analysis_res}

Table \ref{tab:data_set1} presents the analysis dataset for the atopic dermatitis case study, interventions and key covariates. 

\begin{figure}[!htb]
    \centering
    \begin{subfigure}{.49\textwidth}
    \centering
    \includegraphics[width=\linewidth]{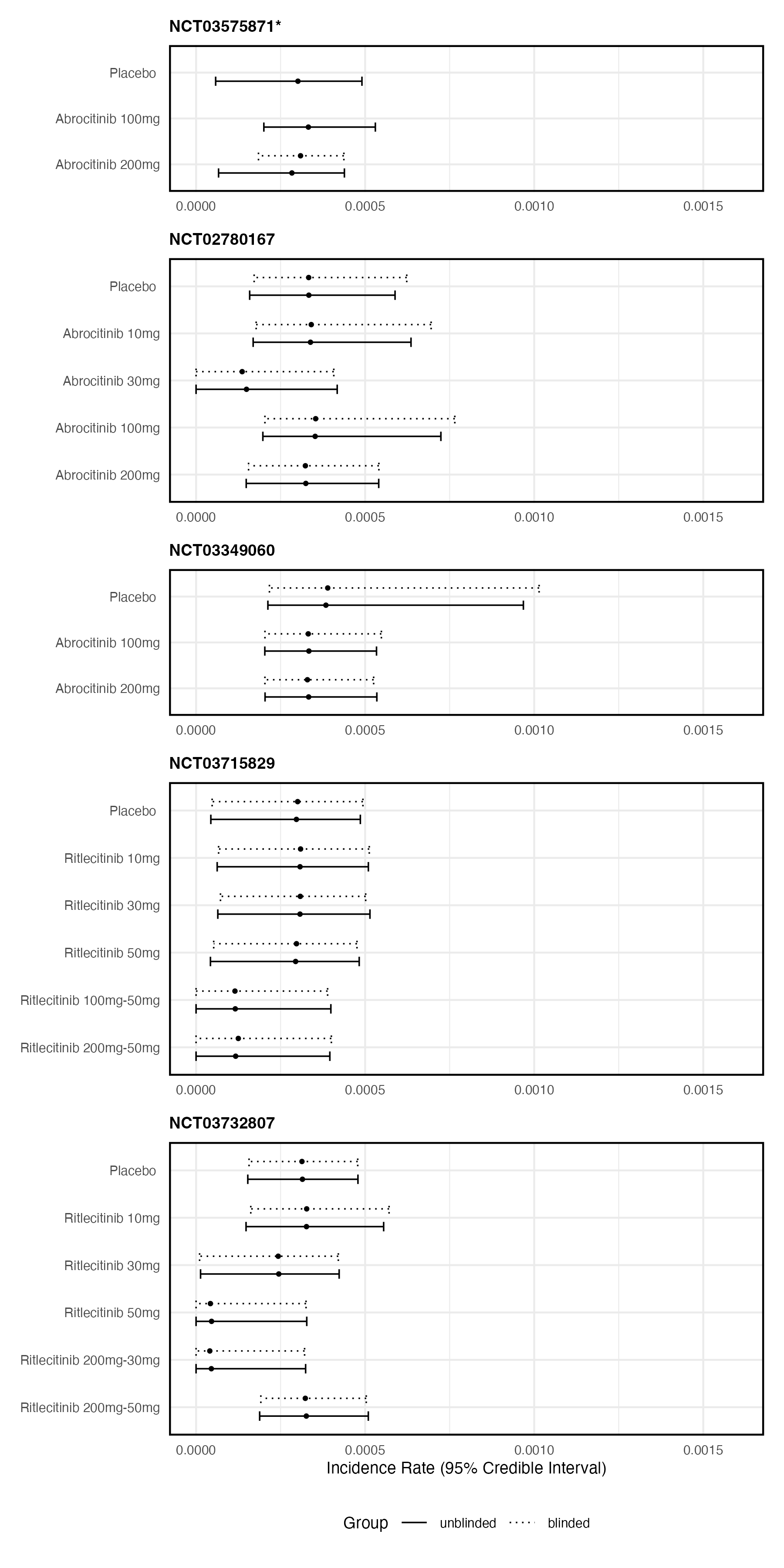}
    \caption{PPMx with total mass $M = 10$}\label{fig:analysis_PPMx_M10}
    \end{subfigure}
    \begin{subfigure}{.49\textwidth}
    \centering
    \includegraphics[width=\linewidth]{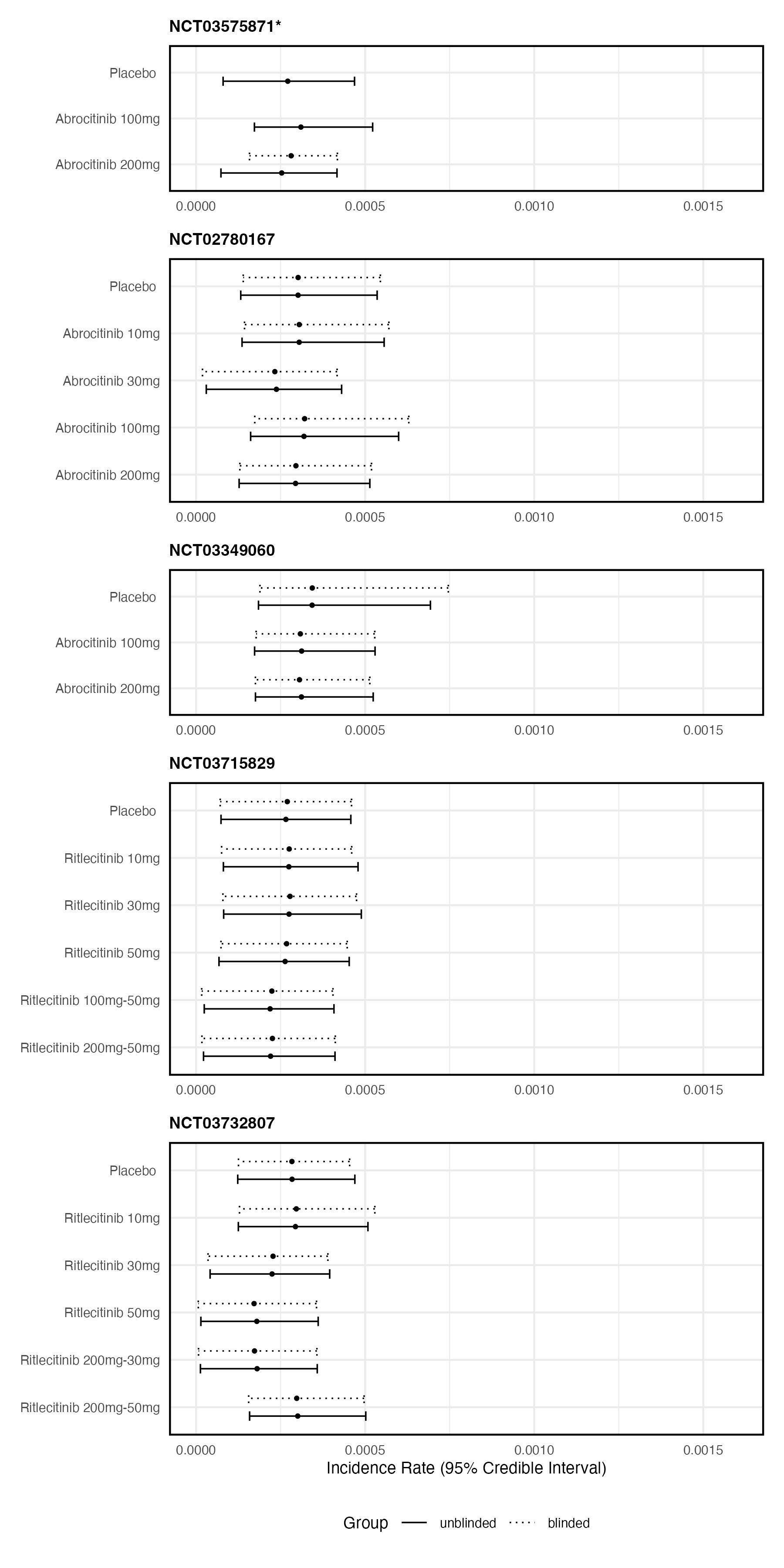} 
    \caption{PPMx with hyperprior $\text{Gamma}(0.001,0.001)$ }\label{fig:analysis_PPMx_001}
    \end{subfigure}
    \caption{Forest plot of posterior mean AE incidence rates by unit, with 95\% credible intervals, for the atopic dermatitis case study. (a) PPMx with total mass $M  = 10$ in the cohesion function $c(S_k)$. (b) PPMx with hyperprior $\text{Gamma}(0.001,0.001)$ on both hyperparameters $a$ and $b$.}
    \label{fig:analysis_res_additional}
\end{figure}

\begin{figure}[!htb]
    \centering
    \includegraphics[width=0.6\linewidth]{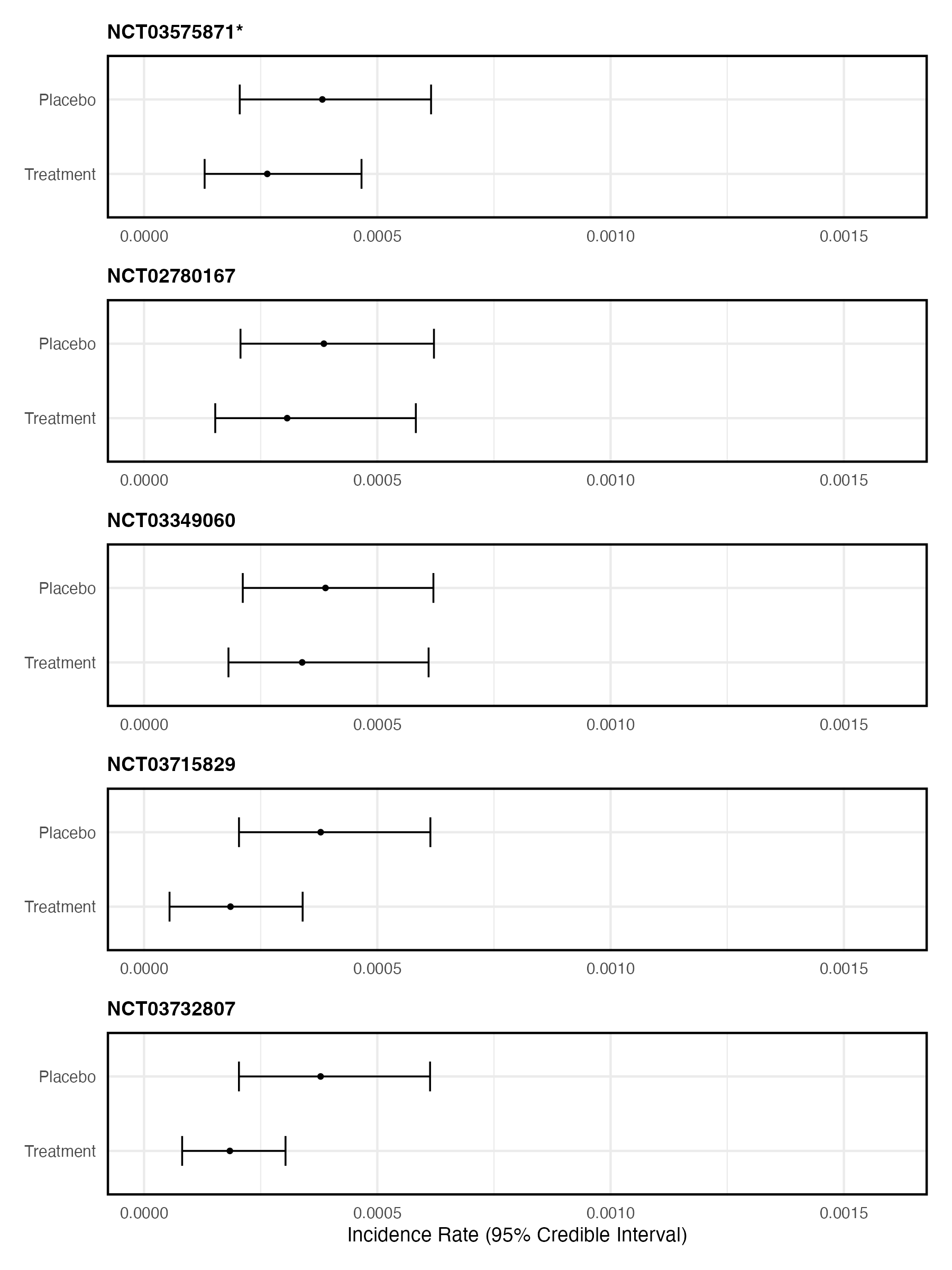}
    \caption{Forest plot of posterior mean AE incidence rates by treatment arm, with 95\% credible intervals, for the atopic dermatitis case study using the proposed model in \citet{cai2010meta} with the analysis settings described in Section~\ref{sec:cai2010}.}
    \label{fig:analysis_res_cai}
\end{figure}

Figures~\ref{fig:analysis_res_additional} and \ref{fig:analysis_res_cai} present posterior summaries of AE incidence rates for the atopic dermatitis case study under alternative modeling approaches. Figure~\ref{fig:analysis_res_additional} shows forest plots based on the PPMx model with different prior specifications: (a) using a fixed total mass parameter $M=10$, and (b) using a hyperprior  $\text{Gamma}(0.001,0.001)$ on both hyperparameters  $a$ and $b$. Figure~\ref{fig:analysis_res_cai} displays results from the model proposed in \citet{cai2010meta}, implemented with the analysis settings detailed in Section~\ref{sec:cai2010}. Each plot displays posterior mean incidence rates with 95\% credible intervals by unit or treatment arm.

\begin{sidewaystable}[!htb]
    \scriptsize
    \centering
    \caption{\scriptsize Analysis dataset for the atopic dermatitis case study. Values of empirical IRs are scaled by \(10^{-4}\) and rounded to two decimal places for readability. The unit labeled ``200mg-50mg'' represents a regimen of ritlecitinib, consisting of a 200 mg daily loading dose followed by a 50 mg daily maintenance dose. The exact follow-up times for NCT03715829 and NCT03732807 are not available. To approximate patient-time for these studies, we use the number of patients multiplied by the average follow-up time derived from the other studies in the dataset.}
    \label{tab:data_set1}
    \begin{tabular}{llllllrrrrr}
\hline
        NCT &           Phase &                     Condition & Intervention &       Dose &                  stgAge &   n &  male &     t &  SAE & Empirical IR \\
\hline
NCT03575871 &         phase 3 &             Atopic Dermatitis &      Placebo &         & CHILD,ADULT,OLDER\_ADULT &  78 &    47 &  5257 &    1 &     1.90 \\
NCT03575871 &         phase 3 &             Atopic Dermatitis &  Abrocitinib &      100mg & CHILD,ADULT,OLDER\_ADULT & 158 &    94 & 12419 &    5 &     4.03 \\
NCT03575871 &         phase 3 &             Atopic Dermatitis &  Abrocitinib &      200mg & CHILD,ADULT,OLDER\_ADULT & 155 &    88 & 12617 &    2 &     1.59 \\
NCT02780167 &         phase 2 &             Atopic Dermatitis &      Placebo &         &       ADULT,OLDER\_ADULT &  56 &    21 &  4589 &    2 &     4.36 \\
NCT02780167 &         phase 2 &             Atopic Dermatitis &  Abrocitinib &       10mg &       ADULT,OLDER\_ADULT &  49 &    21 &  4056 &    2 &     4.93 \\
NCT02780167 &         phase 2 &             Atopic Dermatitis &  Abrocitinib &       30mg &       ADULT,OLDER\_ADULT &  51 &    22 &  4412 &    0 &     0 \\
NCT02780167 &         phase 2 &             Atopic Dermatitis &  Abrocitinib &      100mg &       ADULT,OLDER\_ADULT &  56 &    31 &  5188 &    3 &     5.78 \\
NCT02780167 &         phase 2 &             Atopic Dermatitis &  Abrocitinib &      200mg &       ADULT,OLDER\_ADULT &  55 &    28 &  5602 &    2 &     3.57 \\
NCT03349060 &         phase 3 &             Atopic Dermatitis &      Placebo &         & CHILD,ADULT,OLDER\_ADULT &  77 &    49 &  5713 &    4 &     7.00 \\
NCT03349060 &         phase 3 &             Atopic Dermatitis &  Abrocitinib &      100mg & CHILD,ADULT,OLDER\_ADULT & 156 &    90 & 12277 &    5 &     4.07 \\
NCT03349060 &         phase 3 &             Atopic Dermatitis &  Abrocitinib &      200mg & CHILD,ADULT,OLDER\_ADULT & 154 &    81 & 12243 &    5 &     4.08 \\
NCT03715829 &         phase 2 & Active Non-segmental Vitiligo &      Placebo &         &       ADULT,OLDER\_ADULT &  66 &    40 &  5329 &    1 &     1.88 \\
NCT03715829 &         phase 2 & Active Non-segmental Vitiligo & Ritlecitinib & 200mg-50mg &       ADULT,OLDER\_ADULT &  65 &    30 &  5248 &    0 &     0 \\
NCT03715829 &         phase 2 & Active Non-segmental Vitiligo & Ritlecitinib & 100mg-50mg &       ADULT,OLDER\_ADULT &  67 &    31 &  5410 &    0 &     0 \\
NCT03715829 &         phase 2 & Active Non-segmental Vitiligo & Ritlecitinib &       50mg &       ADULT,OLDER\_ADULT &  67 &    39 &  5410 &    1 &     1.85 \\
NCT03715829 &         phase 2 & Active Non-segmental Vitiligo & Ritlecitinib &       30mg &       ADULT,OLDER\_ADULT &  50 &    28 &  4037 &    1 &     2.48 \\
NCT03715829 &         phase 2 & Active Non-segmental Vitiligo & Ritlecitinib &       10mg &       ADULT,OLDER\_ADULT &  49 &    25 &  3956 &    1 &     2.53 \\
NCT03732807 & phase 2,phase 3 &               Alopecia Areata &      Placebo &         & CHILD,ADULT,OLDER\_ADULT & 131 &    86 & 10577 &    3 &     2.84 \\
NCT03732807 & phase 2,phase 3 &               Alopecia Areata & Ritlecitinib &       10mg & CHILD,ADULT,OLDER\_ADULT &  62 &    43 &  5006 &    2 &     3.99 \\
NCT03732807 & phase 2,phase 3 &               Alopecia Areata & Ritlecitinib &       30mg & CHILD,ADULT,OLDER\_ADULT & 132 &    80 & 10658 &    1 &     0.94 \\
NCT03732807 & phase 2,phase 3 &               Alopecia Areata & Ritlecitinib &       50mg & CHILD,ADULT,OLDER\_ADULT & 130 &    71 & 10496 &    0 &     0 \\
NCT03732807 & phase 2,phase 3 &               Alopecia Areata & Ritlecitinib & 200mg-30mg & CHILD,ADULT,OLDER\_ADULT & 130 &    85 & 10496 &    0 &     0 \\
NCT03732807 & phase 2,phase 3 &               Alopecia Areata & Ritlecitinib & 200mg-50mg & CHILD,ADULT,OLDER\_ADULT & 132 &    81 & 10658 &    4 &     3.75 \\
\hline
\end{tabular}
\end{sidewaystable}

\clearpage\newpage

\clearpage\newpage
\section{Simulated Scenarios} \label{sec:sim_sc}
Table \ref{tab:sim_sc} summarizes the simulation scenarios used in Section~\ref{sec:sim}, including scaled hypothetical incidence rates and simplified treatment unit labels for similarity measurement.

\begin{table}[htbp]
  \centering
  \scriptsize
  \caption{Simulation scenarios used in Section \ref{sec:sim}. Values of hypothetical IRs are scaled by \(10^{-4}\) and rounded to two decimal places for readability. The unit labeled ``200mg-50mg'' represents a regimen of ritlecitinib, consisting of a 200 mg daily loading dose followed by a 50 mg daily maintenance dose. For simplicity, we treat this unit as 200 mg and measure similarity accordingly. This approach is applied to other similar units as well.}
  \smallskip
        \begin{tabular}{lllrrrrr}
    \hline
    \multicolumn{1}{c}{\multirow{2}[4]{*}{NCT}} & \multicolumn{1}{c}{\multirow{2}[4]{*}{Intervention}} & \multicolumn{1}{c}{\multirow{2}[4]{*}{Dose}} & \multicolumn{1}{c}{\multirow{2}[4]{*}{n}} & \multicolumn{1}{c}{\multirow{2}[4]{*}{t}} & \multicolumn{3}{c}{Hypothetical IR} \\
\cline{6-8}          &       &       &       &       & \multicolumn{1}{l}{Scenario 0} & \multicolumn{1}{l}{Scenario 1} & \multicolumn{1}{l}{Scenario 2} \\
    \hline
    NCT03575871 & Placebo &       & 78    & 5257  & 3     & 6     & 3 \\
    NCT03575871 & Abrocitinib & 100mg & 158   & 12419 & 3     & 5.53  & 6.91 \\
    NCT03575871 & Abrocitinib & 200mg & 155   & 12617 & 3     & 6.36  & 7.95 \\
    NCT02780167 & Placebo &       & 56    & 4589  & 3     & 3     & 3 \\
    NCT02780167 & Abrocitinib & 10mg  & 49    & 4056  & 3     & 2.76  & 3.45 \\
    NCT02780167 & Abrocitinib & 30mg  & 51    & 4412  & 3     & 4.08  & 5.1 \\
    NCT02780167 & Abrocitinib & 100mg & 56    & 5188  & 3     & 5.53  & 6.91 \\
    NCT02780167 & Abrocitinib & 200mg & 55    & 5602  & 3     & 6.36  & 7.95 \\
    NCT03349060 & Placebo &       & 77    & 5713  & 3     & 3     & 3 \\
    NCT03349060 & Abrocitinib & 100mg & 156   & 12277 & 3     & 5.53  & 6.91 \\
    NCT03349060 & Abrocitinib & 200mg & 154   & 12243 & 3     & 6.36  & 7.95 \\
    NCT03715829 & Placebo &       & 66    & 5329  & 3     & 3     & 3 \\
    NCT03715829 & Ritlecitinib & 200mg-50mg & 65    & 5248  & 3     & 3     & 3 \\
    NCT03715829 & Ritlecitinib & 100mg-50mg & 67    & 5410  & 3     & 3     & 3 \\
    NCT03715829 & Ritlecitinib & 50mg  & 67    & 5410  & 3     & 3     & 3 \\
    NCT03715829 & Ritlecitinib & 30mg  & 50    & 4037  & 3     & 3     & 3 \\
    NCT03715829 & Ritlecitinib & 10mg  & 49    & 3956  & 3     & 3     & 3 \\
    NCT03732807 & Placebo &       & 131   & 10577 & 3     & 3     & 3 \\
    NCT03732807 & Ritlecitinib & 10mg  & 62    & 5006  & 3     & 3     & 3 \\
    NCT03732807 & Ritlecitinib & 30mg  & 132   & 10658 & 3     & 3     & 3 \\
    NCT03732807 & Ritlecitinib & 50mg  & 130   & 10496 & 3     & 3     & 3 \\
    NCT03732807 & Ritlecitinib & 200mg-30mg & 130   & 10496 & 3     & 3     & 3 \\
    NCT03732807 & Ritlecitinib & 200mg-50mg & 132   & 10658 & 3     & 3     & 3 \\
    \hline
    \end{tabular}%
  \label{tab:sim_sc}%
\end{table}%

\clearpage\newpage
\section{Model and Decision Problem of \citet{cai2010meta}} \label{sec:cai2010}

Let $j = 1, \ldots, J$ index available studies, 
comprising the current study ($j = 1$) and $J - 1$ historical studies. 
Each study has two treatment arms: a control arm ($i = 0$) and a treated arm ($i = 1$). For each study $j$ and treatment arm $i$, let $Y_{ij}$ denote the number of adverse events observed during the total exposure time $t_{ij}$. We assume $Y_{ij} \sim \text{Poisson}(t_{ij}\theta_{ij})$, where $\theta_{ij}$ is the incidence rate.

\citet{cai2010meta} proposed a random relative risk model of the form
\begin{equation}
    \theta_{ij} = \xi_j \exp(\tau_j A_{ij}),
\end{equation}
where $A_{ij}$ is a binary indicator ($A_{ij}=1$ for the treated arm, 0 for the control arm). In this framework, the random effect $\xi_j$ is assumed to follow a $\text{Gamma}(\alpha,\beta)$ distribution, and the log relative risk $\tau_j$ is assumed to follow a normal distribution $N(\mu, \sigma^2)$. We place non-informative priors on all the parameters in the model. Specifically, we let
\begin{align*}
    \mu \sim \text{Uniform}(-10^6,10^6), \quad \sigma \sim \text{Uniform}(0,10^6), \\
    \alpha \sim \text{Uniform}(0,10^6), \quad \beta \sim \text{Uniform}(0,10^6).
\end{align*}

For the desired comparison, we augment the inference model with  rules based on decision boundaries of \eqref{eq:pi_unblind} and \eqref{eq:pi_trt_ctrl}, using the following two probabilities and thresholds

\begin{equation}
    \pi_2 \equiv \Pr\!\left(\theta_{11} > \frac{1}{2J - 1}\left[\theta_{10} + \sum_{j=2}^I\sum_{i=0,1}\theta_{ij}\right] + \delta \quad \Bigg| \quad \mathbf{y},  \mathbf{t}\right) > \lambda_2,
\end{equation}
\begin{equation}
    \pi_3 \equiv \Pr\!\left(\theta_{11} > \theta_{10} + \delta \quad \Bigg| \quad \mathbf{y},  \mathbf{t}\right) > \lambda_3.
\end{equation}


\clearpage\newpage
\section{Model and Decision Problem of \citet{mukhopadhyay2018bayesian}} \label{sec:mukhopadhyay2018}
Let $j = 1, \ldots, J$ index available studies, 
comprising the current study ($j = 1$) and $J - 1$ historical studies. 
Each study has two treatment arms: a control arm ($i = 0$) and a treated arm ($i = 1$). For each study $j$ and treatment arm $i$, let $Y_{ij}$ denote the number of adverse events observed during the total exposure time $t_{ij}$. 


\citet{mukhopadhyay2018bayesian} only consider inference under blinding. Instead of having separate data for the treated and control arms, only the combined event count $Y_{\cdot 1} = Y_{01} + Y_{11}$, the total exposure time $t_{\cdot 1} = t_{01} + t_{11}$, the ratio of exposure times, $k = t_{11} / t_{01}$ are assumed to be known. 
They assume the known background rate
They model the total event count as 
\begin{equation}
    Y_{\cdot 1} \sim \text{Poisson}\left(\theta_0^* \cdot t_{\cdot 1} \cdot \frac{rk + 1}{k + 1}\right)
\end{equation}
where $r$ can be also represented by $r = \frac{p}{k(1-p)}$, with $p$ denoting the probability that an adverse event (AE) originates from the treated arm. \citet{mukhopadhyay2018bayesian} recommend using a non-informative prior on $p$, such as Beta(0.5, 0.5) or U(0, 1). In this manuscript, we adopt U(0, 1).

They comment
The background rate, denoted as $\theta_0^*$, can be obtained, for example, by Bayesian meta-analysis to combine information from multiple historical studies. Specifically, we implement the following Bayesian meta-analysis to estimate $\theta_0^*$.
We assume, for $j = 2, \ldots, J$,
\begin{equation}
    \begin{aligned}
        Y_{0j} \sim \text{Poisson}\left(\theta_{0j} t_{0j}\right) \\
        \log(\theta_{0j}) \sim N(\mu, \sigma^2)
    \end{aligned}
\end{equation}
Priors are placed on the hyperparameters $\mu$ and $\sigma$ as follows:
\[
\mu \sim \text{Uniform}(-1000,1000), \quad \sigma \sim \text{Uniform}(0,100)
\]
Therefore, the background rate $\theta_0^*$ is then computed as the average of the posterior means of the incidence rates $\theta_{0j}$ from the control arms of the historical trials:
\begin{equation}
    \ths_0 = \frac{1}{J- 1} \sum_{j = 2}^J E[\theta_{0j} \mid Y_{0j}, t_{0j}, j = 2, \ldots, J]
\end{equation}

For the comparison of Table \ref{tab:sim_res},  we recommend unblinding if 
\begin{equation}
    \pi_1 \equiv \Pr\left( r \ths_0  > \ths_0 + \delta \mid  Y_{\cdot 1}, t_{\cdot 1}\right) > \lambda_1,
\end{equation}

\end{appendices}

\end{document}